\documentclass{article}
\usepackage[T1]{fontenc} 
\usepackage[utf8]{inputenc} 
\usepackage{ismir,amsmath,cite,url}
\usepackage{graphicx}
\usepackage{xcolor}
\usepackage{glossaries}
\usepackage{xspace}
\usepackage{amssymb}
\usepackage{booktabs}
\usepackage[bookmarks=false,hidelinks]{hyperref}
\usepackage{microtype}

\usepackage{lineno}
\def\ie{i.e.,\xspace}
\def\eg{e.g.,\xspace}

\newcommand{\ActDataset}{\mathcal{X}}

\newcommand*\rot{\rotatebox{70}}

\newacronym{nmf}{NMF}{Non-negative Matrix Factorisation}
\newacronym{ntd}{NTD}{Non-negative Tucker Decomposition}
\newacronym{cav}{CAV}{Concept Activation Vector}
\newacronym{tcav}{TCAV}{Testing with Concept Activation Vectors}
\newacronym{maestro}{MAESTRO}{MIDI and Audio Edited for Synchronous TRacks and Organization}
\newacronym{svm}{SVM}{Support Vector Machine}
\newacronym{mir}{MIR}{Music Information Retrieval}
\newacronym{cnn}{CNN}{Convolutional Neural Network}

\title{Concept-Based Techniques for ``Musicologist-friendly'' Explanations in a Deep Music Classifier}

\multauthor
{Francesco Foscarin$^{1*}\thanks{* Equal contribution.}$ \hspace{1cm} Katharina Hoedt$^{1*\footnotemark[1]}$ \hspace{1cm} Verena Praher$^{1*\footnotemark[1]}$ } { \bfseries{Arthur Flexer$^1$ \hspace{1cm} Gerhard Widmer$^{1,2}$}\\
 $^1$ Institute of Computational Perception, Johannes Kepler University Linz, Austria\\
$^2$ LIT AI Lab, Linz Institute of Technology, Austria\\
{\tt\small \{firstname\}.\{lastname\}@jku.at}
}

\def\authorname{F. Foscarin, K. Hoedt, V. Praher, A. Flexer, and G. Widmer}

\begin{document}

\maketitle
\begin{abstract}
Current approaches for explaining deep learning systems applied to musical data provide results in a low-level feature space, \eg by highlighting potentially relevant time-frequency bins in a spectrogram or time-pitch bins in a piano roll. This can be difficult to understand, particularly for musicologists without technical knowledge. To address this issue, we focus on more human-friendly explanations based on high-level musical concepts. Our research targets trained systems (post-hoc explanations) and explores two approaches: a supervised one, where the user can define a musical concept and test if it is relevant to the system; and an unsupervised one, where musical excerpts containing relevant concepts are automatically selected and given to the user for interpretation. We demonstrate both techniques on an existing symbolic composer classification system, showcase their potential, and highlight their intrinsic limitations.

\end{abstract}


\section{Introduction}\label{sec:introduction}



The mass adoption of deep learning methods in recent years has increased interest in the field of \textit{explainability},\footnote{Considered synonymous to the term \textit{interpretability} in this paper.} \ie the study of techniques that generate a human-understandable explanation of a model's decision~\cite{miller2019explanation}. 
As deep learning models are usually not intrinsically interpretable, techniques that can be applied to trained models (\ie \textit{post-hoc} methods) are of great interest. 
The resulting explanations cannot only reveal potential issues of the system itself and the data it uses, but can also provide insights into the problem we are targeting, thus helping us to gain knowledge about it~\cite{molnar2020interpretable}.

Higher-level musical tasks such as chord transcription or composer classification may require explanations that can only be understood by persons with advanced musical expertise.
However, the explanation techniques for musical systems that have been proposed in recent years~\cite{mishra2017local, mishra2020reliable,Haunschmid2020MML, Haunschmid2020arxiv, melchiorre2021ecir} are \textit{feature-based}, \ie the explanation is given in terms of the input features the system considers. Typical input features for musical systems, \eg spectrograms or piano roll representations, are high-dimensional, and, since the importance of a single time frequency / pitch bin does not convey much meaningful interpretation, feature-based explanations can be hard to understand.\footnote{Moreover, their truthfulness has recently been debated~\cite{cr1,cr2,cr3,praher2021_veracity}.}
Musicologists are therefore often unable to analyse the results of trained systems, let alone contribute to the development of new learning models.
This motivates research on techniques that provide explanations that are as similar as possible to those that a human music domain expert would naturally use. 

\textit{Concept-based} explanations offer an interesting direction. They were first explored by Kim et al.~\cite{kim2018tcav} and later developed in several works (\eg by Chen et al.~\cite{chen2020_conceptwhitening}) for image systems, which also use high-dimensional input features. 
Instead of producing feature-level descriptors, the explanation is based on human-understandable concepts. 
For example, \cite{kim2018tcav} tests whether the concept of ``stripes'' would increase the probability that an image classifier labels an image as ``zebra''. Music can also be described with \textit{musical concepts}; terms such as ``diatonic sequence'', ``alberti bass'', ``difficult-to-play music'', ``orchestral music'',  ``rubato'', ``shuffle drum beat'', ``funky bass line'', etc. are used to describe pieces or specific elements in a piece. 

In this paper, we explore two concept-based techniques to explain deep learning systems that deal with musical data: one supervised and one unsupervised.
The first (see Section~\ref{sec:supervised}) is based on \gls*{tcav} \cite{kim2018tcav}: the user defines a concept by providing examples and interrogates the system to find out if the concept is relevant or not for its decision.
This can be applied to any kind of neural network and, even more generally, to any system that has a hidden layer for which we can compute directional derivatives. 
The second technique, described in Section~\ref{sec:unsupervised}, is an adaptation of \cite{zhang2021invertible} and works in an unsupervised fashion, where the most relevant concepts are automatically produced and given to the user for interpretation. 
Each concept is presented in the form of a set of musical excerpts. 
This approach requires networks whose hidden layers contain only non-negative values and have a spatial correlation with the input data, two conditions that are satisfied by most \glspl*{cnn}. 

Our contributions are: 
the first application of concept-based post-hoc approaches to musical data, in particular, we target the composer classification system of Kim et al.~\cite{composer_class} that uses piano roll representations of piano MIDI files as input; the definition and creation of musical concept datasets; a dedicated visualisation of unsupervised concepts for symbolic music data; and, finally, the exploration of the \gls{ntd} for the factorisation of hidden layers. Our code and data are available on Github.\footnote{\url{https://github.com/CPJKU/composer_concept}}
%


%


\section{Related Work}
\label{sec:Related Work}
Recent work in the field of \gls*{mir} that focus on explainability for deep models consists mainly of feature-based post hoc methods (\eg \cite{mishra2017local,mishra2020reliable, Haunschmid2020MML, melchiorre2021ecir}). Chowdhury et al. introduce pre-defined ``mid-level features'' that could be considered concepts as intermediate targets in a two-level prediction model~\cite{chowdhury2019TowardsExplainableEmotion}. Related to this, approaches that consider instrinsic as opposed to post-hoc methods gain increasing attention in the audio domain as well (e.g.,~\cite{zinemanas2021interpretable,zinemanas2021dcase,ren2022prototype}).
However, no prior studies have examined post-hoc concept-based explainability techniques on systems that work with musical data. To provide a technical context for our work, we focus on related approaches that work on audio or image data.

A recent approach~\cite{asokan2022interpretability} applies concept-based techniques to multimodal data (video, audio, and text) to explain an emotion classifier for video sequences of human conversations. 
For the audio signal, they only test the concept of ``voice pitch'', \ie the averaged fundamental frequency of the speaker's voice. 
In another related study, Parekh et al.~\cite{parekh2022listen} learn a codebook of sounds (\eg alarm sound) from input audio through \gls*{nmf}, which is then used to obtain hidden network layer representations that indicate time activations of these pre-learnt components.
This has some similarities with our unsupervised approach, as we also make use of non-negative factorisation techniques to disentangle concepts. 
However, while~\cite{parekh2022listen} performs the factorisation on the input and propagates the results to a hidden layer, we factorise the hidden layer activations and project the results back to the input data.
The approach of \cite{parekh2022listen} is promising if we assume that the underlying reason for a system decision can be extracted directly from the input with unsupervised separation approaches. However, since this might not always be the case, we factorise the layer activations to exploit the non-linear feature extraction a network does internally
to obtain more meaningful explanations.

Our work uses techniques and results originally proposed for the image domain. The work of Kim et al.~\cite{kim2018tcav} provides the basis for the supervised explanation, although the creation of concept data sets is more challenging for music. We base the unsupervised explanation on the work of Zhang et al.~\cite{zhang2021invertible}, but propose a dedicated visualisation of piece excerpts and test different solutions for the tensor factorisation step by employing the \gls*{ntd}.



\section{Experimental Setup}
This section details the type of data and the system that we use to demonstrate our explainability techniques. 

{\bf Data:} We use MIDI representations of piano performances from the MAESTRO v2.0.0 dataset~\cite{hawthorne2018enabling}. As proposed by Kim et al.~\cite{composer_class}, we pre-select data by composers with at least 16 pieces and remove files with more than one composer (\eg Schubert/Liszt, ``Der Mueller und der Bach''), so that pieces of 13 different composers remain (see Table~\ref{tab:supervised_results}). We randomly split the resulting 667 pieces in a training (462 pieces) and validation (205 pieces) set. 
For each piece, we sample 90 excerpts of 20 seconds randomly across time and different performances of the same piece (if available). 

{\bf Composer Classifier:} In this work, we investigate the composer classification system proposed by Kim et al.~\cite{composer_class}. For more recent systems, the code was not available~\cite{Kong2020LargeScaleMC} or we were unable to reproduce their results~\cite{yangcomposer}.
During preprocessing, we transform MIDI excerpts into piano roll representations with a 50 ms time step, \ie a matrix $88 \times 400$,
which is used as input to a ResNet-50~\cite{He2016Resnet}. Kim et al.~\cite{composer_class} use an additional channel with onset information, which we omit because it does not improve the performance of our system. As proposed by \cite{composer_class}, we train the network with Stochastic Gradient Descent with momentum (factor 0.9), L2 weight regularisation (factor 0.0001), and cross-entropy loss function. The initial learning rate is set to $0.01$, and scheduled with cosine annealing \cite{Loshchilov2017cosineannealing}. 
Our attempt to retrain the system results in a F1 score of 0.93 compared to 0.83 in the original work~\cite{composer_class}. The accuracy of our system is 0.93.
The difference in performance could be attributed to a problem during preprocessing in the original code, which reduced the resolution of the piano roll.


\section{Supervised concept-based explanations}
\label{sec:supervised}
In this section, we use \gls*{tcav}~\cite{kim2018tcav} to build a supervised concept-based explainer. We manually define musical concepts and interrogate a music classifier to find out how much a concept influences the results of the classifier.

\subsection{Musical Concepts}
Musical concepts describe the characteristics of a certain group of notes and are identified by musicologists with a specific name or with a small sentence (\eg ``staccato'', ``rubato'', 
``melody with jumps'', etc.). To define a musical concept in a way that can be used within our system, we construct \textit{concept datasets}, \ie sets of pieces that have one specific musical concept in common.
In this paper, we build three different concept datasets, each consisting of 30 musical excerpts of \textasciitilde 25 seconds. Ideally, the bigger and more diverse the concept datasets is, the lower is the probability that it will also represent other unwanted concepts.

The first dataset describes the \textit{``alberti bass''}: an accompaniment pattern first used during the classical period, where notes of chords are horizontally distributed in the left hand part~\cite{rosen1997classical}.
For this dataset a semi-professional pianist composed \textasciitilde 25 second excerpts that contain this pattern, while trying to vary as much as possible other musical elements (\eg key, tempo, content of the right hand).

The second concept is \textit{``difficult-to-play music''}. A dataset of difficult musical excerpts was collected using the ranking produced by the musical score publisher G. Henle.\footnote{\url{https://www.henle.de/us/about-us/levels-of-difficulty-piano/}} Excerpts were sampled from difficult pieces available in the MAESTRO dataset among different composers to avoid introducing biases toward some of them.

The third concept is \textit{``contrapuntal texture''}, which denotes piano pieces composed of multiple monophonic voices that behave as separate instruments. This style is mostly present in pieces by some Baroque composers (e.g., Bach, Telemann, Händel, Buxtehude). For this dataset, we sampled Bach fugue performances, ensuring that they were not used during training the targeted composer classifier.

In addition to these three concept datasets, in this paper we use a collection of 10 different \textit{random datasets}, which are built by randomly sampling 20 second excerpts from the MAESTRO dataset.


\subsection{CAVs and Conceptual Sensitivity}
\label{sec:cavs}
\begin{figure}
 \centerline{
 \includegraphics[width=0.9\columnwidth]{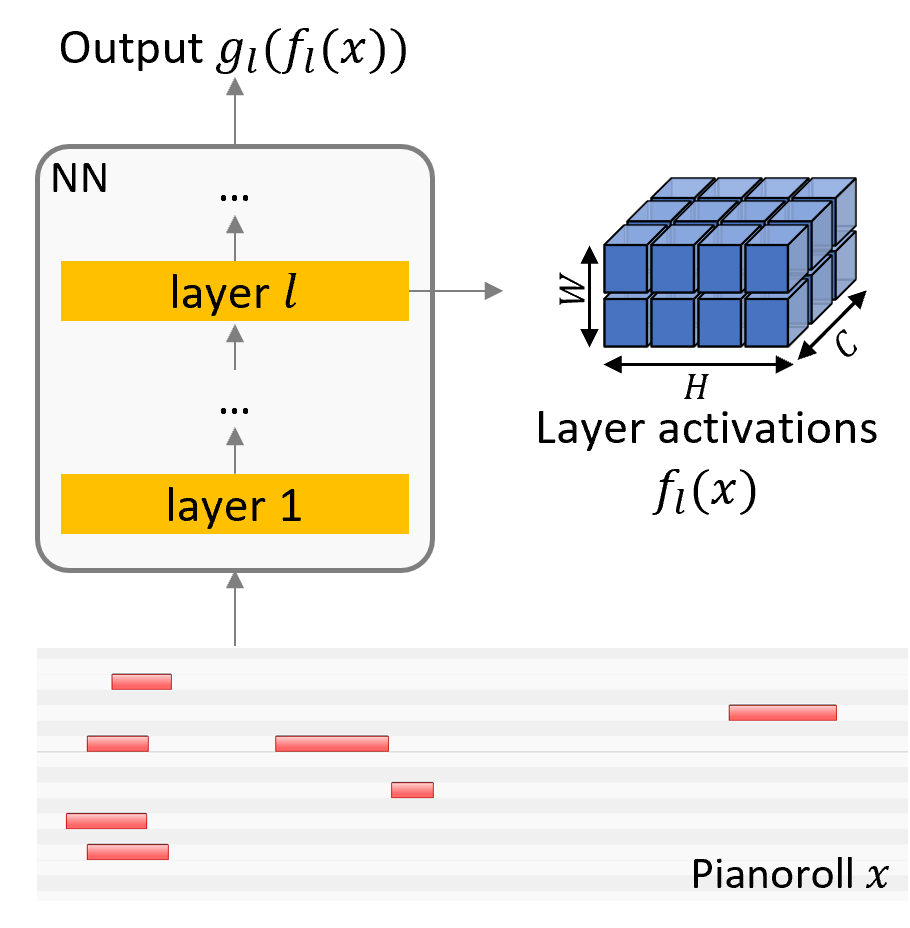}}
 \caption{Layer $l$ activations for one piece excerpt. $H,W$ and $C$ are the horizontal, vertical, and channel size. The function encoded by the neural network is represented in two parts: $f_l$ from the NN input to the layer output, and $g_l$ from the layer output to the NN output.}
 \label{fig:layer_act}
\end{figure}

A \textit{\gls*{cav}}~\cite{kim2018tcav} $\mathbf{v}^k_l$ represents a concept $k$ in the output space of a neural network layer $l$. 
To compute it, we need the corresponding concept dataset (containing \eg pieces with alberti bass) and a random dataset~\cite{kim2018tcav}. For a specific network layer $l$, we compute the layer activations (\ie the output of the layer) for every piano roll $\mathbf{x}$ in the concept dataset, as well as the random dataset (see Figure~\ref{fig:layer_act}). 
These activations can be seen as points in a $(H \times W \times C)$-dimensional space, where $H,W,C$ are the horizontal, vertical, and channel size in the layer activations tensor.
We train a binary linear classifier (\eg \gls*{svm} or logistic regression) that separates the layer activations of the concept pieces from those of the random pieces. The vector of coefficients of this binary classifier, \ie the vector orthogonal to the classification boundary, is the \gls*{cav} $\mathbf{v}^k_l$~\cite{kim2018tcav}.

To measure whether a concept $k$ is relevant for a piece being classified as a certain composer $o$, we use the \textit{conceptual sensitivity} $S_{k,o,l}$ of the system~\cite{kim2018tcav}, \ie the directional derivative of the prediction in the direction of the \gls*{cav},
\begin{equation}
    S_{k,o,l} = \nabla g_{l,o}(f_l(\mathbf{x})) \cdot \mathbf{v}^k_l.
\end{equation}

Here, $g_{l,o}$ transforms the activation vector $f_l(\mathbf{x})$ to the logit for the output class $o$, that is, it represents the remaining computations after a layer $l$ up to the output of the system.
Intuitively, $S$ is a scalar that measures how much the output logits change if we perturb the layer activations in the direction of the \gls*{cav}. Positive values mean that a concept $k$ encourages the classification of $\mathbf{x}$ as class $o$.

The conceptual sensitivity is a local explanation, \ie it explains how a system behaves for a specific input. We produce a global explanation, the \gls*{tcav} score, that no longer depends on a specific input, by taking multiple pieces that belong to one class and computing the ratio of pieces for which $S$ is positive~\cite{kim2018tcav}.

\subsection{Experiments and Results}
\label{subsec:Supervised Experiments and Results}
To investigate \gls*{tcav}, we first compute a \gls*{cav} for every one of our proposed concepts ``alberti bass'', ``difficult-to-play music'', and ``contrapuntal texture''.\footnote{Using \url{https://captum.ai/api/concept.html}}
As a linear classifier that separates the activations of the concept samples from those of the random samples, we use a linear \gls*{svm}. This approach requires inputs of the same dimension, so we crop or pad all concepts to 20 seconds length (also used during training).
Cropping is done by selecting the middle 20 seconds of a MIDI performance; padding adds silence until the appropriate length is reached. 

\begin{table*}[!ht]
    \centering
    \begin{tabular}{ l c c c c c c c c c c c c c } 
     & \rot{Bach} & \rot{Scarlatti} & \rot{Haydn} & \rot{Mozart} & \rot{Beethoven} & \rot{Schubert} & \rot{Chopin} & \rot{Schumann} & \rot{Liszt} & \rot{Brahms} & \rot{Debussy} & \rot{Scriabin} & \rot{Rachmn.} \\
     \midrule
     alberti bass        & + &  & + & + & + &   & - &   & - & + & - & - & - \\
     diff.-to-play music & - &  & - & - &   &   &   & + & + &   &   & + & + \\
     contrapuntal texture   & + &  &   & + &   & - & - &   & - &   & - &   &   \\
    \end{tabular}
    \caption{Summary of TCAV scores for three concepts and the penultimate layer of a composer classifier. ``+'' indicate a positive influence of a concept on the classification of a composer, ``-'' negative influence. Empty cells show results that fail significance testing, \ie the concept does not consistently en-/decourage the classification of a certain composer.
    }
    \label{tab:supervised_results}
\end{table*}

In the next step, we examine the conceptual sensitivities of all the validation data and compute the \gls*{tcav} score for all pieces by the same composer, \ie the relative amount of positive conceptual sensitivities over the pieces. 
We again need to ensure inputs have the same length as our concepts, so we split every piece into non-overlapping 20 second segments and use all of these for subsequent computations. 
Although we can compute the \gls*{tcav} score for any layer of the composer classifier, for brevity we show results of the penultimate layer subsequently, expecting this layer to encode the highest level features, similar to the image domain~\cite{zhou2014}.
To compute \gls*{tcav} scores, we perform ten runs with ten random datasets \cite{kim2018tcav}, and run a two-sided t-test and a Bonferroni correction for all concepts and composers to validate our experiments. We use a significance threshold of $\alpha = 0.05 / 13$ (correcting for 13 hypothesis tests).

In our experiments, the \gls*{svm} differentiating between concepts and random data has an accuracy greater than 0.9 for all concepts (in most cases, even 1). This means that the activations of the penultimate layer for the concept and the random samples are linearly separable, \ie the CAV we produce represents the concept it is targeting.
The TCAV score results are summarised in Table~\ref{tab:supervised_results}. All cells with ``+'' or ``-'' show results that pass our statistical significance test, and the remaining (empty) cells show results that fail.
The symbol ``+'' means that a particular concept appears important for the classification of the corresponding composer (\ie average \gls*{tcav} score $> 0.5$); ``-'' that a concept discourages the classification of an input as a certain composer (\ie average \gls*{tcav} score $< 0.5$). 

Table~\ref{tab:supervised_results} shows both results that we would expect (\eg alberti bass being relevant for Mozart, contrapuntal texture for Bach), and a few results which seem counter-intuitive (\eg the model relying on alberti bass for Bach or Brahms -- although one can find examples of such structures also in their works). 
It is possible that our model mixes the proposed concepts with other confounding ones, \eg all pieces in the Alberti Bass dataset have also a quite simple harmonic structure.
In general, there is no proof that the model \textit{understands} our proposed concepts similarly to how a human listener would, yet we can make some interesting observations.
Remarkably, even an extremely abstract concept such as ``difficult-to-play music'' might be grasped by our model: Liszt, Scriabin, Rachmaninoff are clear candidates for this attribute. The same applies to the ``contrapuntal texture". Cases with negative impact \mbox{(``-'')} should probably be interpreted with care: the fact that the classifier did not consider these concepts relevant for the classification does not necessarily mean that they are not present in the pieces. This could apply in particular to the subtle concept of contrapuntal texture.
Also interesting is the case of Scarlatti, who is very much an outsider in classical music, style-wise (``a freakish if not downright incorrect composer'' \cite{kirkpatrick1983domenico}) and could not be associated with any of our concepts.


\section{Unsupervised Concept-based Explanations}
\label{sec:unsupervised}

The supervised approach requires the user to pre-define concepts. This is very time-consuming, and the user could have to try a potentially infinite number of concepts if the network works differently than expected.
In this section, we discuss an unsupervised approach instead: we build an explainer that identifies the relevant concepts and presents pieces where this concept is maximally activated (and some where the concept is not present). The musical expertise of the user is then used to translate these example pieces into a musical concept (with or without a name).

For the unsupervised approach described below, we introduce two limitations on the target neural network: we assume it to be convolutional and to use a non-negative activation function~\cite{nwankpa2018activation} (e.g., ReLU).

\subsection{Tensor Factorisation for CAV Extraction}
Consider a set of layer activations $\ActDataset = \{f_l(\mathbf{x}_1),\cdots,f_l(\mathbf{x}_N)\}$ generated from multiple pieces $\mathbf{x}_1,\ldots,\mathbf{x}_N$. Layer activations that are close together (in terms of Euclidean distance) correspond to perceptually similar inputs~\cite{zhang2018unreasonable} and therefore might describe similar concepts within the inputs. We could cluster similar activations and consider the pieces that generate these activations as examples of the same concept~\cite{ghorbani2019towards}.

This works best if only one concept is present in a piece excerpt. However, we expect an excerpt to contain a number of different musical concepts, \eg an alberti bass and a legato melody, both following a certain chord progression. Since these concepts can be shared across the same notes, we need a way to disentangle their effects on the layer activations. Due to the restriction on the type of activations (only non-negative) that we introduced in this section, we can use the \gls*{ntd} for this objective (see Section~\ref{sec:ntd}). 

\subsection{Channel CAVs}
\begin{figure}
 \centerline{
 \includegraphics[width=0.99\columnwidth]{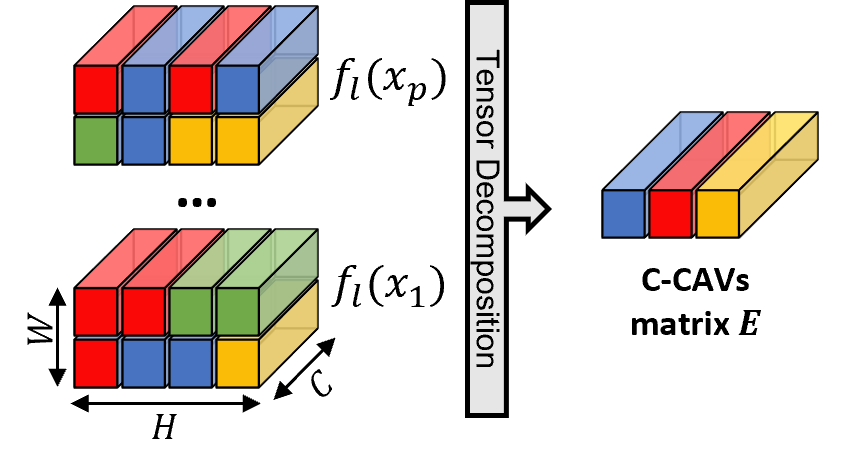}}
 \caption{Left: Channel-mode tubes that result from fixing indices of activations in the $W$ and $H$ dimension of the activation space. Right: C-CAVs extraction from the layer activations of multiple pieces. Each channel tube is decomposed as a weighted sum of $C'$ C-CAVs.}
 \label{fig:channel_extraction}
\end{figure}

Given layer activations $f_l(\mathbf{x})$, let us consider their \textit{channel-mode tubes}~\cite{tensor_decomposition}, \ie the vectors obtained by fixing an index $h$ and $w$ for the horizontal and vertical dimension (see left-hand side of Figure~\ref{fig:channel_extraction}).
In the case of \glspl*{cnn}, we can consider each of these vectors as a different representation of the same piece with a different receptive field \cite{luo2016understanding}. As proposed in \cite{zhang2021invertible}, we can analyse channel-mode tubes $\in \mathbb{R}^{C}$ instead of full layer activations $\in \mathbb{R}^{H \times W \times C}$. This increases the amount of data and reduces the dimension of each data point by a factor of $H \times W$, therefore, we expect that the tensor decomposition that we will run on these data will achieve better results.
Since every channel-mode tube represents a piece, we can compute \glspl*{cav} in this restricted channel space $\mathbb{R}^{C}$, and refer to them as \textit{Channel-CAVs} (or C-CAVs). We can again compute the conceptual sensitivities for such C-CAVs, as explained in Section~\ref{sec:cavs}.

We compute C-CAVs by starting from a dataset of pieces (segmented in 20-second excerpts) of the composers we want to explain (any number of composers can be considered).
We input each excerpt into the trained system and produce activations for a certain layer $l$ (see Figure~\ref{fig:layer_act}). The set of all activations can be seen as a tensor $\ActDataset \in \mathbb{R}^{N \times H \times W \times C}$, where $N$ is the number of piece excerpts and $H, W$ and $C$ are the frequency, time, and channel size of the layer activation tensor. We then apply a \gls*{ntd} to $\ActDataset$ to obtain a set of C-CAVs.

Moving to a channel-based formulation also permits us to highlight in which part of a piece a certain concept is present: since layer activations have a spatial correlation with input data~\cite{dai2015convolutional} we can project the position $h,w$ of each C-CAV onto the input piano roll. This creates a \textit{concept presence} heatmap showing the presence of a C-CAV~\cite{zhang2021invertible} on a piano roll, which can be visualised to improve the user's understanding of the concept. Averaging the values in this heatmap gives a number that expresses how much a concept is activated in a certain excerpt and allows a ranking of the pieces according to their average concept presence.

\subsection{Non-negative Tucker Decomposition}
\label{sec:ntd}
The \gls*{ntd} is a technique to decompose a tensor (\eg $\ActDataset$) into a so-called non-negative \textit{core} tensor $\mathcal{T}$, and multiple \textit{factor} matrices $\mathbf{A} \in \mathbb{R}^{N \times N'}$, $\mathbf{B} \in \mathbb{R}^{H \times H'}$, $\mathbf{D} \in \mathbb{R}^{W \times W'}$ and $\mathbf{E} \in \mathbb{R}^{C \times C'}$ (one for every dimension)~\cite{tensor_decomposition}, such that

\begin{equation}
\label{eq:ntd}
    \ActDataset \approx 
    \sum_{n=1}^{N'} \sum_{h=1}^{H'} \sum_{w=1}^{W'} \sum_{c=1}^{C'} t_{nhwc} \mathbf{a}_n \circ \mathbf{b}_h \circ \mathbf{d}_w \circ \mathbf{e}_c
\end{equation}

Here, the core tensor $\mathcal{T}$ is in $\mathbb{R}^{N' \times H' \times W' \times C'}$, and one of its scalar elements is denoted by $t_{nhwc}$.
The symbol ``$\circ$'' denotes the vector outer product of the column vectors of the (four) factor matrices $\mathbf{a}_n, \mathbf{b}_h, \mathbf{d}_w, \mathbf{e}_c$. The number of columns of the factor matrices (\ie the \gls{ntd} ranks), $N', H', W'$ and $C'$ are hyper-parameters that can be chosen by the user; if we set them to $N' = N$, $H' = H$, etc., an exact reproduction of the original tensor $\ActDataset$ is possible\cite{tensor_decomposition}. 
However, we mostly want to set them at lower values (\ie $N' << N$, $H' << H$, etc.) to decrease the size of the matrices.  

Equation~\ref{eq:ntd} tells us that every channel-mode tube in $\ActDataset$ can be reconstructed as a weighted sum of the columns of matrix $\mathbf{E}$. As previously mentioned, each channel-mode tube represents a piece (in activation space), and each piece contains a sum of multiple concepts as C-CAVs. Then the C-CAVs we are looking for are disentangled as columns in $\mathbf{E}$ (see Figure~\ref{fig:channel_extraction}), and their number is specified by rank C'.





The \gls*{ntd} also allows us to reconstruct an approximation of the original tensor (\ie the original layer activations), and compute the output of the composer classifier by feeding it back into the network. 
We can then compute the ratio of the predictions that remain unchanged after the \gls*{ntd} step (\ie the \textit{fidelity})~\cite{zhang2021invertible} to evaluate its impact on the composer classifier. 
For more details on \gls*{ntd}, we refer to \cite{tensor_decomposition}; in this paper, we use the implementation provided in~\cite{tensorly}, with the Hierarchical non-negative Alternating Least Squares algorithm to update the factor matrices and the Fast Iterative Shrinkage-Thresholding Algorithm to update the core.

\begin{figure*}[!h]
\centerline{
 \includegraphics[width=2\columnwidth]{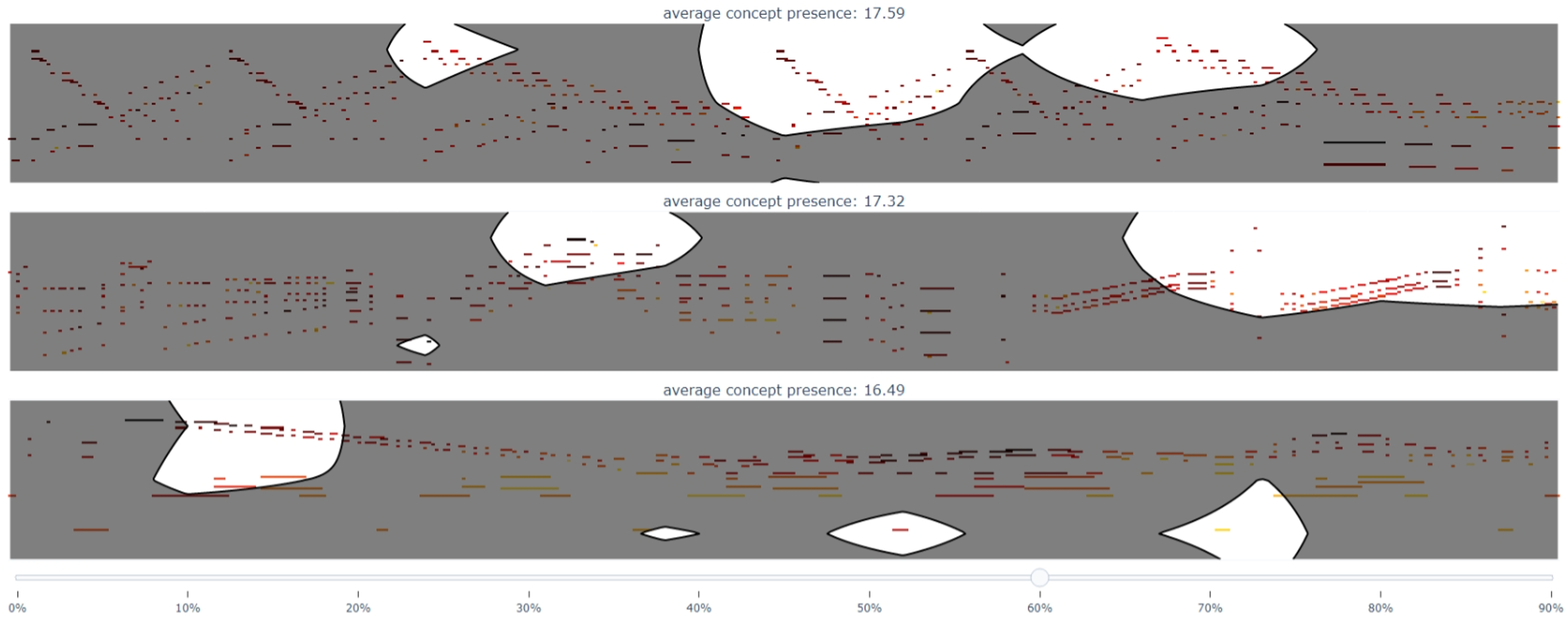}}
 \caption{
Visualisation of one concept (out of 4 produced by 4d NTD) through the masked piano rolls of the three dataset excerpts with the highest average concept presence. The concept heatmap threshold is set to 60\%. This concept has positive conceptual sensitivity for Chopin and negative for Bach, therefore it is useful to distinguish between the two composers.}
 \label{fig:heatmap}
\end{figure*}


\subsection{Experiments and Results}
We compute the unsupervised explanations for the penultimate layer of our composer classification system and test multiple NTD ranks. 
As in \cite{zhang2021invertible}, we present each concept to the user through the five piece excerpts with the highest average concept presence.
The presentation of piece excerpts is more challenging for musical data than for images. 
Although symbolic performances can be visualised with piano rolls, some musical elements (\eg harmonic elements) may be hard to understand in this format. 
We opt for a mixed audio-image visualisation where each excerpt is represented both with a piano roll (with a colour scale for velocity information) and with a listenable MIDI file. 
We create interactive piano roll visualisations (using Plotly~\cite{plotly}) in which the user can zoom in and out to explore different resolution levels. The concept presence heatmap is displayed as a semi-transparent mask over the piano roll. A heatmap with a fixed threshold, as proposed in \cite{zhang2021invertible}, is hard to interpret for our data, so the user is presented with a slider that adapts the heatmap threshold (see Figure~\ref{fig:heatmap}).
We also provide ``contrastive examples'' for each concept, \ie the 5 excerpts where the average concept presence is minimal.
Although our explainer could find relevant concepts starting from a dataset that includes any number of composers, we focus on the results with only two composers. According to psychological studies~\cite{molnar2020interpretable}, explanations are easier to understand when they involve only a small amount of information and when they target contrast cases~\cite{miller2019explanation}, \ie understanding why a composer is selected instead of another is easier than understanding why a composer is selected in general.
For this reason, we also focus on C-CAVs with opposing conceptual sensitivities, \ie negative for one class and positive for the other.
We experimented with three different non-negative factorisation approaches: \gls*{ntd} applied to the 4d matrix (as explained in Section~\ref{sec:ntd}), \gls*{ntd} on a 3d matrix with concatenated horizontal and vertical dimensions, and \gls*{nmf} on a 2d matrix (as proposed by~\cite{zhang2021invertible}) with concatenated horizontal, vertical, and piece dimension.

We found that the target classifier, when only two composers are considered, can be approximated with maximum fidelity by using only 3 to 5 C-CAVs, depending on the composers considered.
For a fixed number of C-CAVs, we found no clear advantage for one of the three factorisation techniques with respect to the fidelity score. \gls*{ntd} allows for a much higher compression of $\ActDataset$, up to 15 times smaller, preserving the same fidelity; but it is also much slower to compute.
From a manual analysis, we see that our unsupervised explainer finds, for each opposing concept, examples of typical composing styles that are useful for discriminating between two composers.

Figure~\ref{fig:heatmap} shows an example of what our model considers a typical Chopin-style pattern that is not present in Bach's music. In musical terms, it might be named ``fast upward or downward movements (in the upper register) in parallel or broken thirds/sixths/octaves".
Since our approach is based on non-negative factorisation techniques, some of their typical problems are also present in our results. For example, our system could produce one C-CAV that comprises what musical experts would typically interpret as two different concepts, or vice versa, produce two C-CAVs, both referring to the same concept. The former might have happened with the four small, seemingly unrelated blobs in the lower registers in the last two piano rolls of Figure~\ref{fig:heatmap}. Furthermore, it is difficult to assign musically meaningful concept names to some C-CAVs, especially those with low average concept presence. 

\section{Conclusion and Future Work}
\label{sec:Conclusion and Future Work}
In this paper, we explored a supervised and an unsupervised approach with the aim of producing explanations of deep musical classifiers interpretable by musicologists. 
In the supervised approach, we define high-level musical concepts (\eg alberti bass) by building concept datasets and interrogate a classifier to find the relevance of a concept for the classifier decisions.
This approach is useful when the user wants to test a specific concept. However, the process of creating a concept dataset can be time-consuming, requires high-level music expertise, and it could be necessary to try many different concepts before finding a relevant one.
A solution to these problems is the unsupervised approach, which selects the relevant concepts by itself. Each concept is presented as a set of piece excerpts where the concept is maximally present. The user can listen to those excerpts and visualise them in a piano roll representation with a heatmap highlighting the concept position.

Future work on the supervised explainer will integrate recent promising results on model non-linearity and stricter hypothesis testing~\cite{pfau2021robust}. 
The unsupervised part will benefit from a formal user-based evaluation by musicologists to see which number of C-CAVs produce the most interpretable musical concepts and if there is agreement on their naming. Sparsity constraints applied to the core tensor and matrices in the \gls*{ntd} may attenuate the non-negative factorisation problems.
While both supervised and unsupervised approaches work on piece excerpts of fixed length, an extension to variable length pieces could enable the study of concepts that span a longer time frame (\eg piece structure).
Moreover, our two approaches could be applied to explain audio classifiers, although this would complicate the visualisation of the concept heatmap for the unsupervised explainer, and could make the creation of concept datasets more challenging.
Finally, dedicated user interfaces enabling to define concepts and visualise results would be helpful for musicologists.

\section{Acknowledgements}
This work was supported by the European Research Council (ERC) under the EU's Horizon 2020 research \& innovation programme, grant agreement No.\ 101019375 (\textit{Whither Music?}), the Austrian Science Fund (FWF, project No. P31988), and the Federal State of Upper Austria (LIT AI Lab).

\bibliography{ISMIRtemplate}

%
%
%
%
%

\end{document}